# Enhancing Phishing Detection through Feature Importance Analysis and Explainable AI: A Comparative Study of CatBoost, XGBoost, and EBM Models

Abdullah Fajar[1,2], Setiadi Yazid[2*], Indra Budi[2]

[1] Telkom University, Bandung Indonesia
[2] University of Indonesia, Depok Indonesia

Corresponding Author Email: setiadi@cs.ui.ac.id

## ABSTRACT

Phishing attacks remain a persistent threat to online security, demanding robust detection methods. This study investigates the use of machine learning to identify phishing URLs, emphasizing the crucial role of feature selection and model interpretability for improved performance. Employing Recursive Feature Elimination, the research pinpointed key features like "length_url," "time_domain_activation" and "Page_rank" as strong indicators of phishing attempts. The study evaluated various algorithms, including CatBoost, XGBoost, and Explainable Boosting Machine, assessing their robustness and scalability. XGBoost emerged as highly efficient in terms of runtime, making it well-suited for large datasets. CatBoost, on the other hand, demonstrated resilience by maintaining high accuracy even with reduced features. To enhance transparency and trustworthiness, Explainable AI techniques, such as SHAP, were employed to provide insights into feature importance. The study's findings highlight that effective feature selection and model interpretability can significantly bolster phishing detection systems, paving the way for more efficient and adaptable defenses against evolving cyber threats.

## 1. INTRODUCTION

Phishing remains a significant cybersecurity threat that targets individuals and organizations by attempting to obtain sensitive information through deceptive means, such as fake websites, emails, or messages. Traditional signature-based detection methods are often inadequate in identifying newly created phishing sites, which constantly evolve to bypass detection. This limitation has prompted the need for machine learning-based approaches that can analyze and predict phishing attempts by examining various features of websites in real time [1], [2]. Machine learning models, when equipped with relevant features, have the potential to effectively predict phishing cases. However, the primary challenge lies in selecting the most relevant features to maximize detection accuracy while maintaining computational efficiency[3], [4].

Feature selection plays a crucial role in building high-performance classification models, as it reduces the dimensionality of the dataset, minimizes overfitting, and enhances model efficiency without compromising accuracy[5]. Various feature selection techniques, such as Information Gain, TreeSHAP, and Principal Components Optimization, have been employed to identify the most significant features[6], [7]. Despite their effectiveness, there is no consensus on the best approach for feature selection in phishing detection, as the impact of each technique varies depending on the dataset and context.

Phishing attacks utilize a range of deceptive techniques, including email spoofing, DNS spoofing, and social engineering, making them challenging to detect and counter. The increasing sophistication of these attacks necessitates the development of robust and adaptive detection methods. Recent studies have shown that machine learning algorithms, such as Random Forest, Naive Bayes, and XGBoost, are effective in detecting phishing websites. However, the performance of these models is highly dependent on the quality and relevance of the features selected from the dataset[8], [9].

The problem statement of this research revolves around the escalating threat of phishing attacks and the inadequacy of traditional detection methods to identify newly created phishing websites, commonly known as zero-day attacks. Traditional signature-based approaches are often ineffective against zero-day phishing attacks, leaving users vulnerable. Therefore, the challenge is to develop machine learning-based methods that can accurately and efficiently detect phishing websites by selecting the most relevant features from large and complex datasets.. This study addresses the following key research questions:

1. How can feature selection methods reduce the number of features while improving the efficiency and accuracy of machine learning models in detecting phishing websites?
2. Which machine learning algorithms perform best in phishing detection when combined with effective feature selection techniques?
3. How can Explainable AI (XAI) methods be used to clearly identify the most influential features in phishing detection and provide a better understanding of their impact on the model's predictions?

These research questions aim to develop robust, efficient, and adaptive phishing detection systems that can operate effectively across various scenarios. This study not only evaluates the performance of feature selection techniques in enhancing the accuracy and efficiency of machine learning models but also leverages XAI methods such as SHAP (SHapley Additive exPlanations) to explain the influence of each feature on the prediction outcomes. The insights gained can guide the development of more transparent and interpretable phishing detection models, ultimately contributing to more secure and reliable cybersecurity solutions

## 2. CONCEPTUAL REVIEW

### 2.1 Feature Selection in Phishing Detection

The features of the datasets that play an important role in the performance of phishing detection models include:
1. **URL-based Features**: These features are derived from the characteristics of the URL itself. Examples include the length of the URL, the presence of an IP address, the use of HTTPS, and the presence of suspicious keywords. URL-based features are crucial because they can be quickly analyzed and are not dependent on the content of the webpage, making them suitable for real-time detection[2], [5]
2. **Content-based Features**: These features involve an in-depth analysis of the webpage content, such as the presence of certain HTML tags, JavaScript functions, and the structure of the webpage. Content-based features help in identifying phishing websites that mimic legitimate ones by analyzing the actual content presented to the user[2], [5].
3. **External-based Features**: These features rely on third-party services such as WHOIS information, search engine indexing, and page rank. They provide additional context about the legitimacy of the website by checking its registration details, popularity, and indexing status. External-based features are useful for cross-verifying the authenticity of a website [5].

Feature selection plays a crucial role in improving phishing detection using machine learning. It helps identify the most relevant features, reducing model complexity and training time while maintaining or enhancing accuracy [1], [3]. Various feature selection techniques have been employed, including correlation-based methods, wrapper techniques, and ranking algorithms like Information Gain and TreeSHAP [5], [8]. Studies have shown that feature selection can significantly improve classification accuracies for algorithms such as Random Forest, Naïve Bayes, and Neural Networks [2], [4]. However, the effectiveness of feature selection methods may vary depending on the dataset and chosen algorithms [9]. While feature selection enhances efficiency and accuracy, it's important to note that some approaches may still struggle with detecting zero-day phishing attacks [7]. Overall, feature selection is essential for developing efficient and effective phishing detection models.

Several gaps are identified in the research on phishing detection using machine learning and feature selection techniques:

1. **Feature Selection Optimization**: [2] highlights the need for effective feature selection techniques to improve classification accuracy. The study compares two feature selection methods, FSOR and FSFM, and finds that optimized feature selection can significantly enhance performance, but there is still room for improvement in reducing processing time and increasing accuracy. [6] proposes a new feature selection framework, HEFS, which shows promising results but still requires further validation and comparison with other datasets to confirm its effectiveness.
2. **Dataset Variability**: Different studies use different datasets, which makes it challenging to compare results directly. For example, [1] and [2] use the UCI Phishing Websites dataset, while [5] uses ISCX-URL-2016 and another public dataset. This variability in datasets can lead to inconsistent performance metrics and hinders the generalizability of findings.
3. **Algorithm Performance**: Different machine learning algorithms are tested across studies, but there is no consensus on the best-performing algorithm. [1], [4] find Random Forest to be the most effective, while [5] reports high performance with XGBoost. This discrepancy indicates a need for more extensive comparative studies to identify the most robust algorithms for phishing detection. [10] [11] both emphasize the importance of comparing different machine learning algorithms to find the most effective one for phishing detection. However, there is a need for more comprehensive comparisons involving a wider range of algorithms and datasets. [12] compares multiple algorithms but suggests that combining existing solutions like blacklisting, whitelisting, and heuristic methods could provide higher security, indicating a gap in integrating multiple approaches effectively.
4. **Hybrid Approaches**: While some studies, like Zamir et al., (2020), explore hybrid models combining multiple algorithms, this approach is not widely adopted. There is potential for further research into hybrid models that leverage the strengths of different algorithms to improve detection accuracy and robustness. [10], [11] both emphasize the importance of comparing different machine learning algorithms to find the most effective one for phishing detection. However, there is a need for more comprehensive comparisons involving a wider range of algorithms and datasets. [12] compares multiple algorithms but suggests that combining existing solutions like blacklisting, whitelisting, and heuristic methods could provide higher security, indicating a gap in integrating multiple approaches effectively.

Addressing these gaps could lead to more effective and universally applicable phishing detection systems

### 2.2 URL and HTML Feature in Web Phishing Detection

Several studies highlight the importance of URL-based features, such as the presence of suspicious characters, domain age, and IP address information. Moreover, incorporating

HTML features, such as hidden text and form action URLs, has been shown to further enhance detection accuracy, thereby underscoring the multifaceted approach necessary for effective phishing detection strategies across diverse datasets[13], [14] Additionally, the integration of machine learning algorithms with these feature sets significantly improves the overall detection capabilities, as evidenced by findings that report accuracy rates exceeding 96% when utilizing a combination of both URL and HTML features in phishing detection efforts [15]. Furthermore, studies indicate that the incorporation of advanced techniques, such as deep learning models, alongside traditional feature sets can lead to even higher detection rates, emphasizing the need for continuous innovation in phishing prevention methods to stay ahead of evolving threats in the digital landscape[13], [14], [15], [16]. This indicates that as phishing techniques become more sophisticated, the application of both established and emerging technologies in feature analysis is paramount in crafting resilient defenses against such attacks [15], [16]. Moreover, the effectiveness of using a blend of machine learning algorithms and comprehensive feature sets reflects the ongoing evolution of strategies required to combat phishing effectively, as recent studies demonstrate that leveraging URL and HTML features together leads to significant improvements in detection performance and reliability, making them a crucial component of robust phishing mitigation systems.

While feature importance analysis offers valuable insights into the factors contributing to phishing detection, it is essential to recognize its limitations and potential drawbacks. Relying solely on feature importance as the primary metric for model evaluation and deployment can lead to an oversimplified understanding of model performance, potentially overlooking other critical aspects such as robustness, generalizability, and adaptability.[17]

### 2.3 Summary of Feature Importance

Feature importance, by its nature, provides a static snapshot of the model's decision-making process, failing to capture the dynamic and context-dependent nature of phishing attacks. Phishing tactics are continuously evolving, and what may be deemed an influential feature today may become obsolete or even adversarial in the future. Over-emphasis on feature importance could result in the development of brittle models that struggle to adapt to emerging threat patterns, undermining their long-term effectiveness in real-world deployment scenarios [18], [19].

Moreover, feature importance analysis may be susceptible to biases, particularly in complex, high-dimensional datasets commonly encountered in phishing detection. Certain features may appear highly influential due to spurious correlations or the model's inability to capture the underlying causal relationships. This could lead to the prioritization of features that are not genuinely indicative of phishing behavior, potentially compromising the model's reliability and contributing to false positives or missed detections.[14], [20]

To address these concerns, a more comprehensive and balanced approach to model evaluation and deployment is essential. While feature importance remains a valuable metric, it should be considered in conjunction with other performance indicators, such as robustness, generalization, and interpretability. By adopting a multi-faceted evaluation framework, researchers and practitioners can develop detection systems that not only excel in identifying known phishing threats but also maintain their effectiveness in the face of evolving attack vectors, fostering trust and confidence among end-users[21], [22].

Feature importance analysis is a crucial component in developing effective phishing detection models. It provides insights into the key factors contributing to accurate identification of phishing attempts, enabling the refinement of feature sets, targeted detection mechanisms, and enhanced user awareness. By understanding the most influential features, researchers and security professionals can focus on the relevant indicators of phishing behavior, improving the overall performance of detection systems and reducing false positives.[16], [23] However, overreliance on feature importance alone can lead to oversimplified models that struggle to adapt to evolving threats. A comprehensive, multi-faceted evaluation framework, incorporating robustness, generalizability, and interpretability, is essential to craft resilient and trustworthy phishing detection solutions. The synthesis of feature importance analysis and advanced techniques, such as XAI, can further enhance the understanding of feature interactions and foster the development of innovative, targeted approaches to combat the dynamic threat of phishing attacks.[24], [25], [26]

In summary, the extraction and analysis of feature importance in phishing detection models serve as a crucial tool in the continuous effort to combat the growing threat of phishing attacks, empowering researchers, security professionals, and end-users with the knowledge necessary to develop and implement more robust and responsive defense mechanisms. Moreover, understanding which features most significantly impact phishing detection allows for targeted improvements in both algorithm design and user education strategies, ultimately leading to a more informed and secure online environment [24], [27]. In this context, a synthesis of both theoretical insights and practical applications is essential to craft an effective response to evolving phishing threats, thereby enhancing the resilience of detection systems and user awareness initiatives..

### 2.4 XAI Technique in Feature Importance Analysis

Applying Explainable Artificial Intelligence (XAI) techniques to enhance cybersecurity, particularly in phishing and malware detection intent to fulfilled lack of explainability in AI-based malware detection systems, which prevents their application in real-world scenarios. Popular XAI methods include SHAP, LIME, LRP, and attention mechanisms for explaining AI-based malware detection systems [28], [29]. For phishing detection, studies have explored using Explainable Boosting Machine [30], glass box models [31], and visual explanations [32], [33]. These approaches aim to improve user awareness, trust, and decision-making in cybersecurity contexts.

However, some research suggests that certain XAI methods may have unintended negative impacts on overall system performance [34]. In malware detection, CNN-based models with LRP have shown promise for Linux systems. Overall, XAI in cybersecurity presents both opportunities for improved defenses and potential vulnerabilities to adversarial attacks[35].

In case phishing detection it is important to identify the most features that influence phishing behavior. Research on feature importance extraction with XAI in phishing detection has shown promising results. Machine learning models using feature selection techniques have achieved high accuracy in detecting phishing websites[3], [36]. Novel approaches like Lorenz Zonoids for feature selection have been proposed to enhance model interpretability[3]. XAI methods have been applied to explain phishing detection results, improving user awareness and trust [30], [33]. Studies have explored various feature extraction and selection techniques, including Chi-Square, Information Gain Ratio, PCA, and LSA, to improve classification performance [37]. Automated feature extraction tools have been developed to identify important characteristics of phishing websites [38]. These advancements in XAI and feature importance extraction contribute to more effective and interpretable phishing detection systems, enhancing cybersecurity efforts [31].

Feature importance is an essential consideration in phishing detection models, as it provides insights into the key factors that contribute to accurate identification of phishing attempts. Understanding the most influential features allows for the development of targeted and efficient detection mechanisms, refinement of feature sets, and enhancement of user awareness and education. By analyzing feature importance, researchers and security professionals can concentrate on the most relevant indicators of phishing behavior, improving the overall performance of detection systems and reducing false positives. The ability to extract and analyze feature importance is a crucial tool in the ongoing effort to combat the evolving threat of phishing attacks, empowering stakeholders with the knowledge to develop more robust and responsive defense mechanisms

## 3. METHODOLOGY

The methodologies to address the research questions involve

1. Datasets collected and load from the UCI Phishing Websites, Kaggle and Mendeley Data are collected with various number of instances and features. A high-quality dataset that collected should be clean, representative, diverse, and well-labeled, with relevant and well-distributed features.
2. Several datasets have an issue regarding imbalance class label distribution. An ideal distribution should be 50% of each class label. This work employs SMOTE technique to overcome this issue.
3. Before Feature Selection process, initial modelling carried out to resume selection of subset features.
4. Feature selection is essential for building efficient, interpretable, and accurate machine learning models. It reduces complexity, enhances performance, prevents overfitting, and makes models easier to explain. By carefully selecting the most relevant features, the model becomes not only faster and more reliable but also more focused on the most important patterns in the data. One of popular approaches such Recursive Feature Elimination employed at his work.
5. Model Training and Evaluation is executed using selected features and several most suitable algorithms to result best performance. The model is evaluated using the test set and appropriate metrics, such as accuracy or F1-score, depending on the task. Models are evaluated based on accuracy, precision, recall, and processing time, with comparative analysis to identify the best-performing methods
6. **Feature Importance Analysis**: Feature importance analysis using Explainable AI (XAI) is a powerful approach for understanding which features contribute the most to a machine learning model's predictions. XAI techniques such as SHAP (Shapley Additive Explanations) or LIME (Local Interpretable Model-agnostic Explanations) allow us to quantify the influence of each feature, either globally across all predictions or locally for specific instances. By analyzing feature importance, we gain insight into how the model makes decisions, identifying key drivers that affect the outcome. This is crucial for domains phishing detection, where understanding the reasoning behind predictions is as important as accuracy. XAI methods ensure transparency, enabling practitioners to detect biases, validate model behavior, and improve trust in machine learning systems by providing interpretable and actionable explanations of the model's decisions.

## 4. RESULT AND ANALYSIS

### 4.1 Dataset Evaluation

Dataset came from various sources and characteristics, such as describe below:

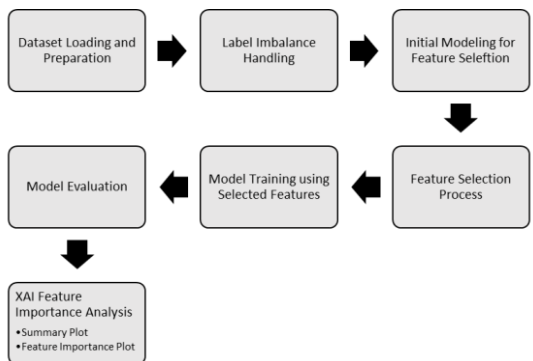

**Figure 1 Methodology**

Table 1 Selected Dataset

| No. | Dataset Name | Source from | URL | Years Created | # Instances | # Features | Phish/Legit Ratio |
|---|---|---|---|---|---|---|---|
| 1 | ds_235795_54 | UCI Datasets | Phiusiil Phishing URL | 2012 | 235,795 | 54 | 43/57 |
| 2 | ds_129K112 | Kaggle | Phishing Domain Detection | 2021 | 129,698 | 112 | 41/59 |
| 3 | ds_100K20 | Mendeley Dataset | Web Page Phishing Dataset | 2020 | 100,000 | 20 | 36/64 |
| 4 | ds_88K112 | Kaggle | Phishing Domain Detection Full | 2021 | 88,647 | 112 | 35/65 |
| 5 | ds_11K89 | Kaggle | Web Page Phishing Detection | 2020 | 11,481 | 89 | 20/80 |
| 6 | ds_11055_32 | Kaggle | Phishing Website Dataset | 2017 | 11,055 | 32 | 44/56 |
| 7 | ds_10K75 | Kaggle | Zieni Dataset | 2024 | 10,000 | 75 | 50/50 |
| 8 | ds_10K50 | Kaggle | Phishing Dataset for ML | 2018 | 10,000 | 50 | 50/50 |
| 9 | ds_10K18 | Kaggle | URL Data for Phishing Website Detection | 2024 | 10,000 | 18 | 50/50 |

Most datasets have a balanced distribution of phishing and legitimate instances, though some are imbalanced. The number of instances varies, with larger datasets not necessarily being more balanced. Newer datasets tend to have more features, and while the creation year doesn't always affect the number of instances, it does correlate with greater feature variation. For optimal phishing detection, it's crucial to consider both the balance and completeness of features in the dataset.

### 4.2 Selected Featured

URL-based features have the most features in the datasets as table1 shown. Then Content-based and External-based features lesser than the URL-based Features. Selected Features may re-groups into several groups for phishing detection as follow:

1. **URL and Domain Structure**: Features like qty_slash_url, length_url, and qty_dot_domain measure URL and domain characteristics, such as the number of slashes or dots, which can indicate suspicious domains.
2. **Directory and File Characteristics**: Features such as directory_length and file_length assess the structure of directories and files in URLs. Long or complex directories can be a sign of phishing.
3. **Time and Certificate Information**: Features like time_response and tls_ssl_certificate evaluate time-related characteristics and SSL certificate details. Short domain activation times or imminent expirations can signal phishing.
4. **IP and Nameserver Details**: Features such as qty_ip_resolved and qty_nameservers measure the number of IP addresses and nameservers. Phishing sites often have unusual or minimal IP and nameserver information.

Based on the frequency analysis of the selected features from various datasets above, the top 10 most frequently appearing features are:

1. **length_url**: Appears 4 times
2. **qty_slash_url**: Appears 2 times
3. **time_domain_activation**: Appears 2 times
4. **web_traffic**: Appears 2 times
5. **qty_redirects**: Appears 2 times
6. **ttl_hostname**: Appears 2 times
7. **qty_mx_servers**: Appears 2 times
8. **qty_nameservers**: Appears 2 times
9. **qty_ip_resolved**: Appears 2 times
10. **time_domain_expiration**: Appears 2 times

The feature length_url is the most frequently appearing, indicating that URL length is a crucial indicator in phishing detection. Other frequently appearing features like qty_slash_url, time_domain_activation, and web_traffic are also important, especially for examining URL characteristics and domain information.

### 4.3 Performance Analysis

This works employs two approaches for modelling such as black box modelling and white box modelling. Black box modelling means that the algorithm has no appropriate explanation how decision made to predict the label and explain what features influence to label prediction. In other hand white box modelling has several capabilities to explain how the model works. Three algorithms such as Random Forest, XGBoost and CatBoost represent as black box modelling approaches and EBM represent as white box modelling approaches. Hereby the performance resume of each models:

1. CatBoost Model. Using this algorithm, A dataset with more features does not always yield the best accuracy. For instance, ds_235795_54 achieved 100% accuracy with fewer features and a smaller dataset. Selecting fewer features can maintain high accuracy, as shown by ds_11055_32 with 96.8% accuracy using only 10 features. Runtime generally increases with the number of features and instances in a dataset, but this does not always correlate directly with model accuracy. Here the result

Table 2 CatBoost Model Result

| No. | Dataset Name | CatBoost Acc | # Selected Feature | # Features | Runtime |
|---|---|---|---|---|---|
| 1 | ds_235795_54 | 100,0% | 4 | 2 | 49 |
| 2 | ds_129K112 | 97,0% | 20 | 20 | 39 |
| 3 | ds_100K20 | 90,0% | 5 | 5 | 40 |
| 4 | ds_88K112 | 97,0% | 20 | 21 | 34 |
| 5 | ds_11K89 | 98,5% | 20 | 26 | 11 |
| 6 | ds_11055_32 | 96,8% | 10 | 10 | 7 |
| 7 | ds_10K75 | 90,0% | 12 | 10 | 8 |
| 8 | ds_10K50 | 99,9% | 14 | 13 | 8 |
| 9 | ds_10K18 | 96,0% | 3 | 3 | 7 |

2. Random Forest. There is no direct correlation between the number of features used in a Random Forest model and its accuracy. For example, the dataset "ds_235795_54" achieved 100% accuracy using only 12 features, while "ds_129K112," despite utilizing 50 features, achieved 99% accuracy. While execution time generally increases with a higher number of features and larger datasets, optimizing feature selection can reduce runtime without significantly impacting accuracy. Larger datasets like "ds_129K112" and "ds_88K112" have longer execution times but generally yield good accuracy. Therefore, carefully selecting the optimal number of features is essential to strike a balance between model accuracy and execution time in Random Forest models..

Table 3 Random Forest Model Result

| No. | Dataset Name | RF | # Features | Runtimes |
|---|---|---|---|---|
| 1 | ds_235795_54 | 100% | 12 | 43.0 |
| 2 | ds_129K112 | 99% | 50 | 36.5 |
| 3 | ds_100K20 | 87% | 4 | 30.0 |
| 4 | ds_88K112 | 97% | 33 | 35.0 |
| 5 | ds_11K89 | 97% | 24 | 4.0 |
| 6 | ds_11055_32 | 94% | 6 | 2.0 |
| 7 | ds_10K75 | 89% | 12 | 2.0 |
| 8 | ds_10K50 | 98% | 12 | 2.0 |
| 9 | ds_10K18 | 96% | 4 | 2.0 |

Compared to CatBoost at the previous table, Overall, CatBoost tends to provide higher accuracy compared to Random Forest on most datasets. For datasets ds_235795_54 and ds_88K112, both models achieve similarly high accuracy, indicating stability on certain datasets. The largest accuracy difference is observed with dataset ds_100K20, where CatBoost outperforms Random Forest by 3%.

3. XGBoost. The relationship between the number of features used in an XGBoost model and its accuracy or execution time is not straightforward. While larger datasets or those with more features might be assumed to yield higher accuracy, this is not always the case. For instance, the dataset "ds_235795_54" achieved 99.6% accuracy using only one feature. Conversely, "ds_10K75," despite utilizing 21 features, only reached 90% accuracy. Execution time is generally lower for smaller datasets or those with fewer features. This is evident in datasets like "ds_10K50" and "ds_11055_32." Notably, smaller datasets like "ds_11K89" can achieve high accuracy (97%) with significantly reduced execution times. These findings underscore that achieving high accuracy with efficient execution times in XGBoost models hinges on identifying and selecting the most relevant features. In cases where a single dominant feature exists, as observed in "ds_235795_54," using fewer features can maintain or even improve model accuracy.

Table 4 XGBoost Model Result

| No. | Dataset Name | XGBoost | XGBosst # Feature | Runtime |
|---|---|---|---|---|
| 1 | ds_235795_54 | 99,6% | 1 | 14.0 |
| 2 | ds_129K112 | 96,0% | 17 | 9.0 |
| 3 | ds_100K20 | 86,0% | 4 | 16.0 |
| 4 | ds_88K112 | 94,3% | 9 | 3.4 |
| 5 | ds_11K89 | 97,0% | 14 | 1.0 |
| 6 | ds_11055_32 | 92,3% | 3 | 0.6 |
| 7 | ds_10K75 | 90,0% | 21 | 0.7 |
| 8 | ds_10K50 | 95,5% | 6 | 0.4 |
| 9 | ds_10K18 | 77,0% | 2 | 0.55 |

When comparing CatBoost, XGBoost, and Random Forest for accuracy and execution time efficiency, CatBoost consistently outperforms the other two algorithms in terms of accuracy across most datasets.

However, when execution time is prioritized, XGBoost emerges as the most efficient, particularly for smaller datasets, demonstrating faster processing times compared to both CatBoost and Random Forest. Here's a concise summary:

a. Prioritize Accuracy: CatBoost is the optimal choice.
b. Prioritize Execution Time: XGBoost is more efficient, especially for smaller datasets.
c. Balance Accuracy and Runtime: CatBoost generally offers a more consistent balance between these two metrics.

In conclusion, CatBoost presents a compelling combination of high accuracy and reasonable runtime efficiency, while XGBoost excels in minimizing execution time.

4. EBM. The number of features used in an EBM model doesn't necessarily dictate its accuracy or execution time. For example, the "ds_235795_54" dataset achieved perfect accuracy (100%) with 17 features. In contrast, the "ds_129K112" dataset, despite utilizing 23 features, only reached 96.7% accuracy. Execution time for EBM models is heavily influenced by dataset size and the number of features. "ds_129K112" exhibited the longest execution time, even though it used only 23 features. Conversely, smaller datasets like "ds_10K18" and "ds_11K89" achieved high accuracy (97% and 97.7%, respectively) with significantly shorter execution times (57 and 34, respectively). These findings suggest that while EBM can achieve high accuracy with a substantial number of features, execution time increases with larger datasets. Therefore, to optimize EBM's efficiency, it's crucial to strike a balance between the number of features used and the desired runtime.

**Table 5 EBM Model Result**

| No. | Dataset Name | EBM | # Features | Runtime |
|---|---|---|---|---|
| 1 | ds_235795_54 | 100,00% | 17 | 677 |
| 2 | ds_129K112 | 96,70% | 23 | 3145 |
| 3 | ds_100K20 | 87,00% | 5 | 190 |
| 4 | ds_88K112 | 96,40% | 39 | 515 |
| 5 | ds_11K89 | 97,70% | 39 | 34 |
| 6 | ds_11055_32 | 94,50% | 13 | 19 |
| 7 | ds_10K75 | 89,00% | 17 | 13 |
| 8 | ds_10K50 | 98,00% | 19 | 13 |
| 9 | ds_10K18 | 97,00% | 4 | 57 |

Overall comparison of all these algorithms may resume as follow:
1. Scalability:
   a. **XGBoost** is the most scalable algorithm due to its efficient handling of large datasets and high-dimensional data, with relatively low execution times.
   b. **CatBoost** is also highly scalable, with the ability to handle large datasets and many features while maintaining high accuracy and reasonable runtime.
   c. **Random Forest** is moderately scalable but can struggle with very large datasets and high feature counts due to increased runtime and memory usage.
   d. **EBM** is the least scalable, as its execution time increases significantly with larger datasets and more features, making it suitable mainly for smaller datasets.
2. **Robustness**
   a. **CatBoost** is the most robust algorithm across the board because: It maintains high and consistent accuracy on various datasets; It scales well with dataset size and number of features: It shows tolerance to feature variability and performs efficiently without a significant runtime increase.
   b. **Random Forest** is a close second in terms of robustness but has more variability in runtime and accuracy.
   c. **XGBoost** is highly efficient in terms of runtime but lacks robustness in accuracy consistency.
   d. **EBM** is less robust due to its sensitivity to runtime and occasional performance drops on larger datasets.

## 4.4 Feature Importance Analysis

Based on performance analysis, the Feature Importance Analysis focuses on two algorithms that have demonstrated the best robustness and scalability: XGBoost and CatBoost. The image is a SHAP (SHapley Additive exPlanations) summary plot, which visualizes the impact of various features on a model's output. Here's a detailed description:

1. **Axes**:
   - **Horizontal Axis**: Represents the SHAP value, indicating the impact of each feature on the model's output. Values to the left (negative) suggest a decrease in the model's prediction, while values to the right (positive) indicate an increase.
   - **Vertical Axis**: Lists the features being analyzed
2. **Color Coding:**
   - The colors range from blue to pink, representing the feature values. Blue indicates lower feature values, while pink indicates higher feature values.
3. **Distribution of SHAP Values:**
   - Each feature has a distribution of SHAP values represented by dots along the horizontal axis. The spread of these dots shows how different values of each feature affect the model's predictions.
4. **Interpretation**:
   - Features show a wider spread, suggesting they have a more significant impact on the model's predictions.
   - The plot helps identify which features are most influential and how their values correlate with the model's output, aiding in understanding the model's decision-making process.

### 4.4.1 CatBoost SHAP Explanation

Here the SHAP Explanation for CatBoost:
1. Dataset 129K112 and 88K112

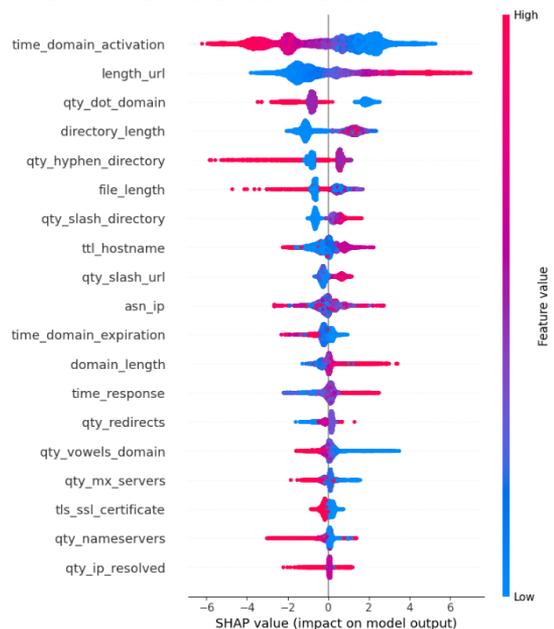

Figure 2 SHAP CatBoost ds_129K112. (Source: Author)

The figure above can be summarized as there are two features that are time_domain_activation and length_url shown most influential and clarity of color distribution to avoid label prediction bias.

2. Dataset ds_235795_54

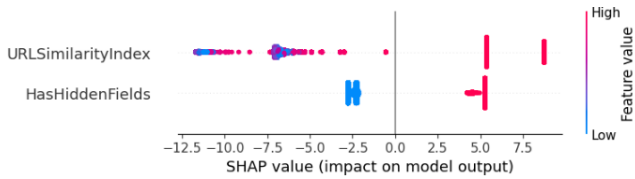

Figure 3 SHAP CatBoost ds_235795_54 (Source: Author)

"URLSimilarityIndex" in the figure above is the most influential feature rather than "HasHiddenFields" but feature value color distribution shown have bias. The color distribution shown is not well distributed between decrease or increase model prediction.

3. Dataset ds_100K20

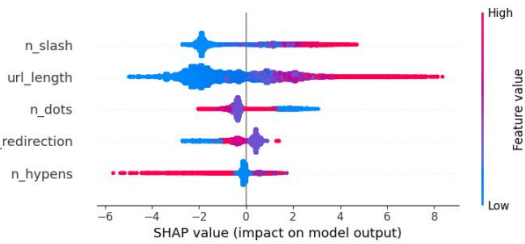

Figure 4 CatBoost SHAP ds_100K20 (Source: Author)

The feature URL_length is the most influential feature but still has potential bias whether higher or lower value in some interval my lead to increase or decrease label prediction.

4. Dataset ds_11K89

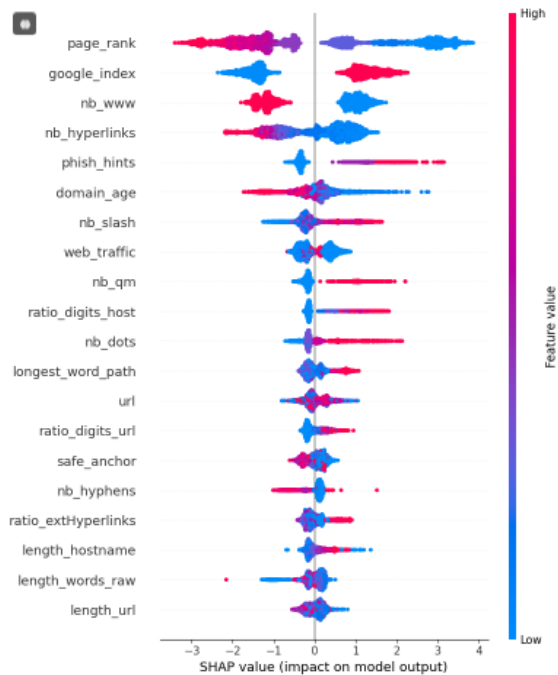

Figure 5 CatBoost SHAP ds_11K89 (Source: Author)

The figure indicates "Page_rank" is the most important feature for predicting the label in the ds_100K20 dataset, it might still have some bias, similar to the "url_length" feature. On the other hand, "google_index" and "nb_www" show a clear separation in their feature value color distribution, suggesting they could be combined with "page_rank" for potentially better label prediction.

5. Dataset ds_11055_32
The dataset result explanation as follow, The features "URL_of_anchor" and "ssl_final_state" have the biggest impact on predicting the label. However, it's not easy to tell from the color distribution of these features whether higher or lower values are more likely to be associated with a positive prediction.

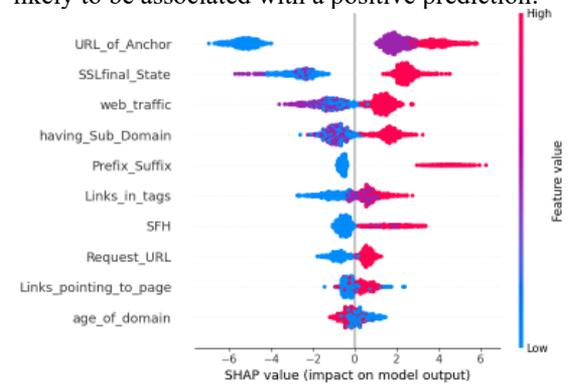

Figure 6 CatBoost SHAP ds_11055_32 (Source: Author)

6. Dataset ds_10K75

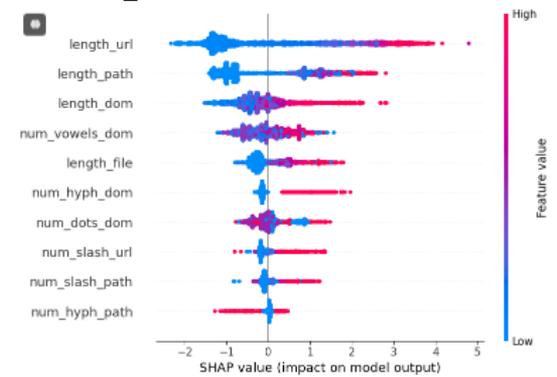

Figure 7 CatBoost SHAP ds_10K75

The following explanation is derived from above figure, which "URL_length" appears to be the most important feature for predicting the label, but there's a potential bias. It's unclear whether a longer or shorter URL, within certain ranges, actually means a phishing attempt is more or less likely.

7. Dataset ds_10K50

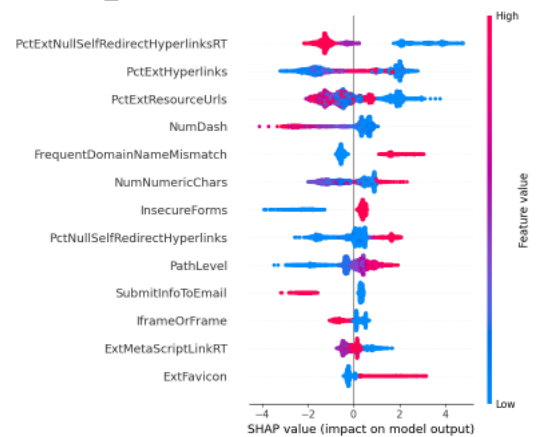

Figure 8 CatBoost SHAP ds_10K50 (Source: Author)

The feature PctExtNullSelfRedirectHyperlinksRT is the most influential feature but still has potential bias whether higher or lower value in some interval my lead to increase or decrease label prediction.

8. Dataset ds_10K18

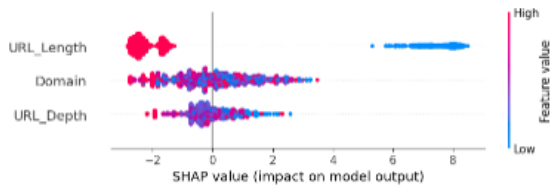

Figure 9 CatBoost SHAP ds_10K18 (Source: Author)

The feature URL_length is the most influential feature but still has potential bias whether higher or lower value in some interval my lead to increase or decrease label prediction.

### 4.4.2 XGBoost SHAP Explanation

Here the SHAP Explanation for XGBoost
1. Dataset ds_129K112 and ds_88K112

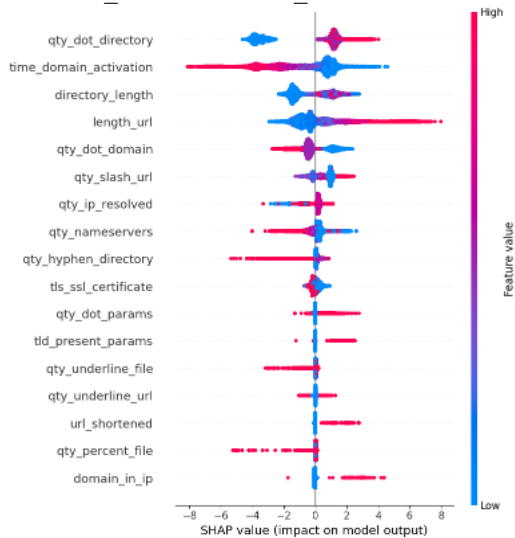

Figure 10 XGBoost SHAP ds_129K112 (Source: Author)

The features tme_domain_activation and length_url is the most influential feature but still has potential bias whether higher or lower value in some interval my lead to increase or decrease label prediction.

2. Dataset ds_235795_32

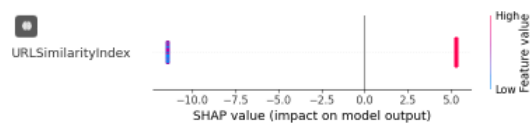

Figure 11 XGBoost SHAP ds_235795_54 (Source: Author)

The figure highlights that a single feature, "URLSimilarlyIndex," was selected during feature selection and is also identified as the most important feature for predicting the label. The author [39] provides further details and explanation regarding this finding.
3. Dataset ds_100K20

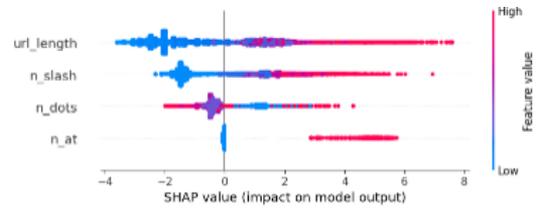

Figure 12 XGBoost SHAP ds_100K20 (Source: Author)

The SHAP figure shows that "URL_length" is the most important feature for predicting the label. However, there seems to be some uncertainty because both longer and shorter URLs, within certain ranges, could potentially lead to an increased or decreased likelihood of phishing.

4. Dataset ds_11K89

As the figure below, although "Page_rank" is the most important factor for predicting the label in the ds_100K20 dataset, it may still have some bias, similar to the "url_length" feature. "google_index" and "nb_www," on the other hand, have a clear color distribution for their feature values, implying that combining them with "page_rank" could improve label prediction accuracy.

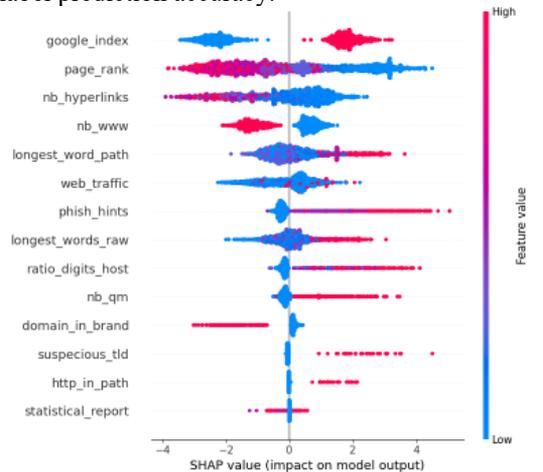

Figure 13 XGBoost SHAP ds_11K89 (Source: Author)

5. Dataset ds_11055_32

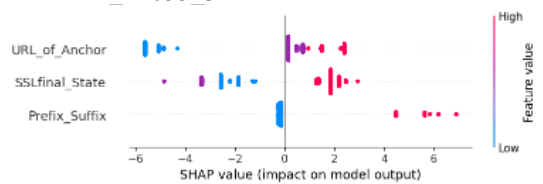

Figure 14 XGBoost SHAP ds_11055_32 (Source: Author)

The SHAP figure visualize and highlights "URL_of_anchor" and "ssl_final_state" as the most important features for predicting the label. However, it's difficult to determine from the color distribution of these features whether higher or lower values are more likely to indicate phishing.

6. Dataset ds_10K75

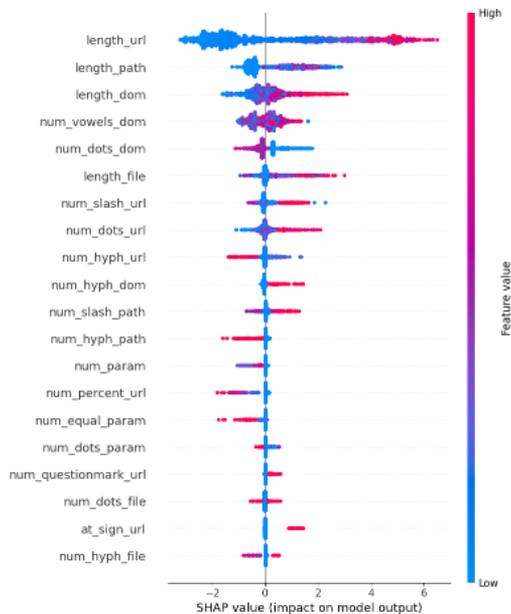
Figure 15 XGBoost SHAP 10K75 (Source: Author)

The feature "URL length" appears to be the most influential for predicting the label. However, there's a potential bias, as both longer and shorter URLs, within certain ranges, could potentially lead to an increased or decreased likelihood of phishing.

7. Dataset ds_10K50
The figura below indicate that the features "PctExtHyperlinks" and "PctExtNullSelfRedirectHyperlinksRT" appear to be the most influential for predicting the label. However, there's a potential bias, as both higher and lower values within certain ranges could potentially lead to an increased or decreased likelihood of phishing.

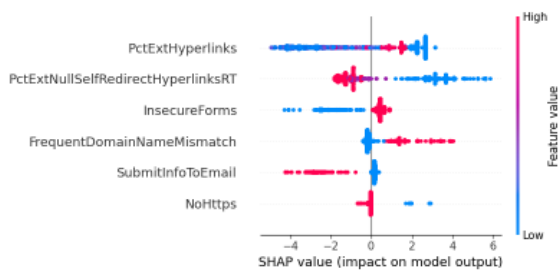
Figure 16 XGBoost SHAP ds_10K50 (Source: Author)

8. Dataset ds__10K18

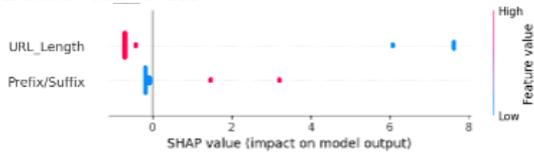
Figure 17 XGBoost SHAP ds_10K18 (Source: Author)

The feature URL_length is the most influential feature but still has potential bias whether higher or lower value in some interval my lead to increase or decrease label prediction

## 5. Discussion And Findings

### 5.1 Discussion

In this part, discussion focuses to answer the following questions that previously state such as:
1. How can feature selection methods reduce the number of features while improving the efficiency and accuracy of machine learning models in detecting phishing websites?
2. Which machine learning algorithms perform best in phishing detection when combined with effective feature selection techniques?
3. How can Explainable AI (XAI) methods be used to clearly identify the most influential features in phishing detection and provide a better understanding of their impact on the model's predictions?

Answer the first question, feature selection method that chosen is RFE or Recursive Feature Elimination. **Recursive Feature Elimination (RFE)** is a feature selection technique used to identify and select the most important features for a machine learning model. The goal of RFE is to select the subset of features that are most relevant and have the highest predictive power, thereby reducing the dimensionality of the dataset and potentially improving model performance.

RFE is particularly useful when dealing with datasets that have many features, as it helps eliminate less important or redundant features, which can reduce overfitting and improve interpretability. RFE works with any estimator that exposes a feature importance attribute or can rank features by their coefficients. Since the dataset used in this works mostly have many features, the range from 18 to 112 Features. Dealing with many features has significantly consume higher computation resources. Here is the explanation about feature reduction by RFE

1. **Impact on Accuracy:**
   - Using Recursive Feature Elimination (RFE) allows most algorithms to maintain accuracy above 95% even when reducing features by up to 75%. This is evident in datasets like ds_235795_54 and ds_129K112.
   - CatBoost and Random Forest exhibit slight accuracy decreases with feature reduction, while XGBoost is more sensitive and shows a significant drop in accuracy with fewer features.

2. **Impact on Runtime:**
   - RFE significantly decreases runtime, particularly for XGBoost and Random Forest. For instance, in dataset ds_129K112, XGBoost's runtime drops to 9 with 17 features, while CatBoost requires 39 runtime units with 20 features.
   - Explainable Boosting Machine (EBM) continues to have higher runtime even after reducing features, indicating that feature reduction has less impact on its runtime.

3. **Feature Reduction Efficiency:**
   - XGBoost shows the highest efficiency in feature reduction. It can reduce from 54 features to just 1, still achieving 99.6% accuracy with the lowest runtime of 14 units.
   - CatBoost and Random Forest also demonstrate good efficiency in feature reduction but need more features than XGBoost to reach similar accuracy levels.

The answer of the second question start from chosen algorithm such as Random Forest, XGBoost, CatBoost and EBM. This work utilizes two modeling approaches: black box and white box modeling. Black box models, such as Random Forest, XGBoost, and CatBoost, lack inherent mechanisms to explain their decision-making process or identify the specific features influencing their predictions. In contrast, white box models, represented in this work by EBM, offer transparency by providing insights into how the model operates and which features contribute to its predictions.

In comparing various machine learning algorithms for web URL phishing detection, CatBoost emerges as the top performer overall. Its impressive accuracy, coupled with its ability to efficiently handle large datasets, makes it a compelling choice. Furthermore, both CatBoost and Random Forest exhibit excellent robustness, maintaining high accuracy even when the number of features is reduced. While EBM also demonstrates good accuracy, it proves less efficient when dealing with larger datasets.

Therefore, when considering both accuracy and efficiency, CatBoost is the recommended algorithm for web URL phishing detection. However, in scenarios where scalability with large datasets is paramount, XGBoost, with its superior runtime efficiency, emerges as the most suitable choice.

The last question to answer is how XAI method can explain clearly which features have most influence in phishing detection and also lead to clear in increasing or decreasing label prediction.

General observations from the SHAP summary plots reveal the influence of individual features on the model's prediction of phishing URLs. The plots illustrate whether a specific feature contributes to an increased or decreased probability of a URL being classified as phishing. Notably, features exhibiting a wider distribution of SHAP values exert a more substantial influence on the model's predictions, highlighting their importance in the decision-making process.

Comparing XGBoost and CatBoost reveals that both algorithms consistently identified "length_url," "time_domain_activation," and "Page_rank" as the most influential features across various datasets. However, a key distinction arises in their feature importance distribution. XGBoost exhibited a tendency to heavily rely on a single dominant feature, such as "URLSimilarityIndex," in certain datasets. In contrast, CatBoost demonstrated a more balanced approach, with multiple features contributing significantly to its predictions. This suggests that CatBoost might offer greater generalizability and robustness by avoiding over-reliance on any single feature.

These features demonstrated significant influence, impacting the model's predictions both positively and negatively. SHAP plots provided valuable visualizations, illustrating how specific feature values contributed to an increased or decreased probability of predicting phishing labels, thereby enhancing the transparency of the models' decision-making processes. However, the analysis also revealed potential biases in certain cases, where higher or lower feature values did not consistently align with expected changes in predictions. This underscores the need for further refinement in feature interpretation to ensure accurate and unbiased phishing detection. Overall, the SHAP analysis effectively pinpointed the most critical features, elucidated their roles in phishing detection, and provided insights into the decision-making processes of both XGBoost and CatBoost models.

## 5.2 Findings

Interesting Findings from this works describe as follow:

1. **Feature Reduction Maintains Accuracy:** Both CatBoost and XGBoost demonstrated the ability to maintain high accuracy even after significant feature reduction using Recursive Feature Elimination. For example, XGBoost achieved 99.6% accuracy on the "ds_235795_54" dataset using only one feature, highlighting that a single dominant feature can be highly predictive in certain cases.

2. **Key Features Remain Consistent:** Across various datasets, features like "length_url," "time_domain_activation," and "Page_rank" consistently emerged as top predictors, indicating their importance in distinguishing phishing URLs. This consistency underscores their significance in the phishing detection process.

3. **Minimal Features Yield Strong Performance:** Remarkably, reducing features by up to 75% using RFE still yielded over 95% of the original model accuracy in many datasets. This suggests that a significant portion of the original features might have limited predictive power and can be omitted without compromising performance.

4. **Single Feature Dominance Observed:** Interestingly, in the "ds_235795_54" dataset, the "URLSimilarityIndex" feature alone proved sufficient for achieving high accuracy. This finding emphasizes that in specific datasets, a single well-selected feature can be dominant, leading to optimal performance and a simpler, faster model.

5. **Potential Bias Requires Attention:** SHAP analysis revealed potential biases in how some features, like "URL_length," impacted predictions. Inconsistent impacts based on feature values, as observed in the "ds_100K20" dataset, suggest potential bias or non-linearity in feature influence, warranting further investigation.

6. **Algorithm Performance Comparison:** XGBoost consistently exhibited the shortest runtime across all datasets, highlighting its efficiency and scalability, especially for large datasets. CatBoost, on the other hand, demonstrated robustness by maintaining accuracy even with significant feature reduction. While EBM achieved high accuracy, its long runtime, particularly on larger datasets, presented a practical limitation.

7. **Summary:** The experiment identified "length_url," "time_domain_activation," and "Page_rank" as crucial features for phishing detection. XGBoost emerged as the most efficient and scalable algorithm, while CatBoost exhibited robustness in accuracy despite feature reduction. EBM, though accurate, faced limitations due to its long runtime. These findings underscore the importance of effective feature selection for optimizing model performance, runtime, and interpretability in phishing detection tasks.

## 6. CONCLUSION AND FUTURE WORKS

**Conclusion:** The works demonstrated that effective feature reduction using Recursive Feature Elimination (RFE) combined with Explainable AI (XAI) methods identified key features such as length_url, time_domain_activation, and Page_rank as the most influential in phishing detection. XGBoost emerged as the most scalable and efficient in terms of runtime, while CatBoost showed high robustness in maintaining accuracy even with reduced features. Although EBM was highly accurate, its long runtimes made it less practical for larger datasets. Overall, the results highlight the importance of optimal feature selection and model interpretability in enhancing phishing detection performance.

**Future Works:**
- **Enhanced Feature Reduction:** Explore advanced techniques like PCA and dynamic feature selection for further optimization.
- **Integration with XAI:** Develop feedback loops between XAI and model training for continuous performance improvement.
- **Real-time Implementation:** Deploy models in live environments to evaluate real-time performance and scalability.
- **New Models and Techniques:** Test other algorithms and anomaly detection methods to handle data imbalance and evolving phishing patterns.
- **Advanced Interpretability:** Utilize additional XAI techniques like LIME and create dashboards for better model transparency.

These future directions will enhance the robustness, scalability, and real-time effectiveness of phishing detection models.

## ACKNOWLEDGMENT

This study was supported by the Faculty of Computer Science University of Indonesia, which provided funding for the publication and dissemination of this research. We express our sincere gratitude for their support in promoting advancements in phishing detection and cybersecurity research.